\pdfoutput=1

\documentclass[manuscript,screen]{acmart}

\AtBeginDocument{%
  \providecommand\BibTeX{{%
    \normalfont B\kern-0.5em{\scshape i\kern-0.25em b}\kern-0.8em\TeX}}}

\setcopyright{custom}
\copyrightyear{2021}
\acmYear{2021}
\acmDOI{https://dbs.ifi.uni-heidelberg.de/files/Team/phausner/publications/Hausner_Gertz_CHI2021.pdf}

\acmConference[CHI 2021]{ACM CHI Virtual Conference on Human Factors in Computing Systems}{May 08--13, 2021}{Yokohama, Japan}
\acmBooktitle{Position Paper at the Workshop "What Can CHI Do About Dark Patterns?" at the CHI Conference on Human Factors in Computing Systems, May 8-13, 2021, Yokohama, Japan}
\acmPrice{15.00}
\acmISBN{XXX-X-XXXX-XXXX-X/XX/XX}

\usepackage{lipsum}
\usepackage{subcaption} 
\usepackage{tikz}
\usepackage[linesnumbered, vlined]{algorithm2e}
\usepackage{cleveref}

\begin{document}

\title{Dark Patterns in the Interaction with Cookie Banners}


\author{Philip Hausner}
\affiliation{%
  \institution{Institute of Computer Science, Heidelberg University}
  \city{Heidelberg}
  \country{Germany}}
\email{hausner@informatik.uni-heidelberg.de}

\author{Michael Gertz}
\affiliation{%
  \institution{Institute of Computer Science, Heidelberg University}
  \city{Heidelberg}
  \country{Germany}}
\email{gertz@informatik.uni-heidelberg.de}

\renewcommand{\shortauthors}{Hausner and Gertz}

\begin{abstract}
Dark patterns are interface designs that nudge users towards behavior that is against their best interests. Since humans are often not even aware that they are influenced by these malicious patterns, research has to identify ways to protect web users against them. 
One approach to this is the automatic detection of dark patterns which enables the development of tools that are able to protect users by proactively warning them in cases where they face a dark pattern.
In this paper, we present ongoing work in the direction of automatic detection of dark patterns, and outline an example to detect malicious patterns within the domain of cookie banners.
\end{abstract}



\maketitle

\section{Introduction}
\label{sec:introduction}
\begin{figure}[t]
  \centering
  \begin{subfigure}{1\linewidth}
    \centering
    \includegraphics[width=0.8\linewidth]{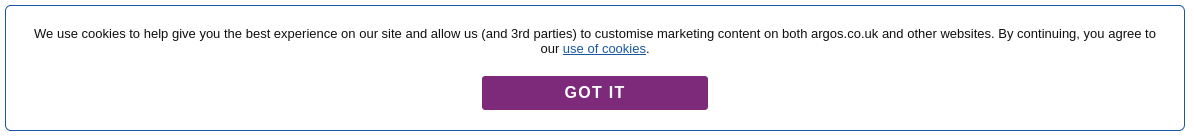}
    \subcaption{}
    \label{fig:cookiebannera}
  \end{subfigure}
  \begin{subfigure}{1\linewidth}
    \centering
    \includegraphics[width=.8\linewidth]{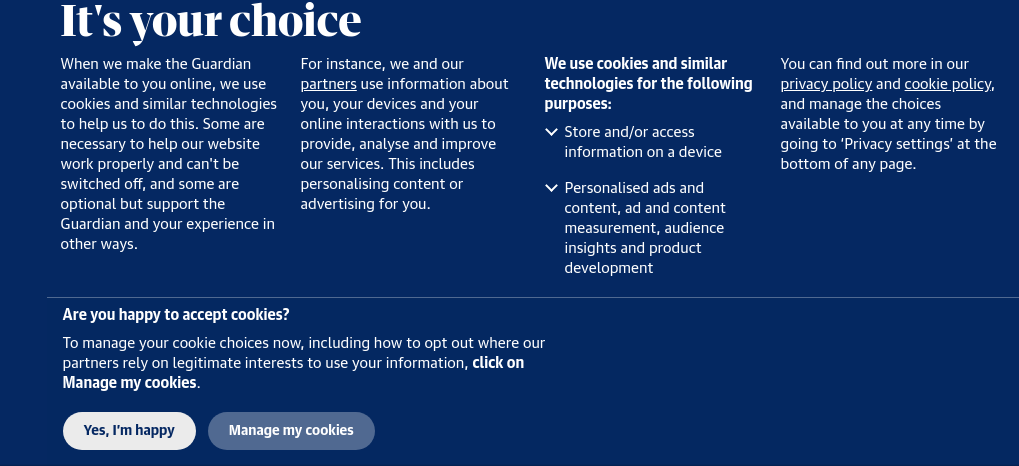}
    \subcaption{}
    \label{fig:cookiebannerb}
  \end{subfigure}
  \caption{Cookie banners found on (a) \url{www.argos.co.uk/} (accessed $2.$ Oct. $2020$), and (b) \url{www.theguardian.com/} (accessed $1.$ Feb. $2021$).}
  \label{fig:cookiebanner}
\end{figure}

Web cookies are small files that store user data during or between browser sessions. Cookies can benefit web site visitors in a variety of ways, e.g., by storing user preferences or remembering a login on a page, and hence, can generally improve user experience. However, recent studies have shown that the large majority of today's web cookies are employed to track users and collect their data with the intent of serving them targeted ads~\cite{Urban2020}. By analyzing this vast amount of data, advertisement companies are able to identify subtle behavioral patterns most users are probably unaware of. While advertisers often claim otherwise, most consumers do not want to be tracked so that companies can serve them "more relevant advertisements": A $2017$ study showed that only about $3\%$ of users are willing to accept web wide tracking cookies\footnote{\url{https://assortedmaterials.com/2017/09/12/new-research-how-many-consent-to-tracking/} (accessed on $3$. February $2021$)}, and research by {\protect\NoHyper\citeauthor{Kulyk2018}\protect\endNoHyper} suggests that most site visitors solely agree to the usage of cookies as a "necessary evil" to access web pages~\cite{Kulyk2018}. Hence, the common practice of setting tracking cookies (not necessarily only for advertisement purposes) for users ignores the privacy concerns of many citizens. To address this issue, the European Union introduced the \textit{General Data Protection Regulation} (GDPR) in $2016$~\cite{GDPR2016}.
As a result, many web pages introduced so-called cookie banners that are frequently implemented as banners at the top or bottom of a page, or as a pop-up that either informs the site visitor about the usage of cookies on the web page, or asks for user consent to set cookies. The complexity of existing banners varies widely, as can be seen in \Cref{fig:cookiebanner}. On one hand, \Cref{fig:cookiebannera} gives the user only the option between the acceptance of cookies or to leave the web site, combined with a short informational text. On the other hand, in \Cref{fig:cookiebannerb} the user is confronted with an information overload, and finally with the choice to accept cookies or to manage the choices. The latter then leads to a submenu presenting the user with a multitude of options to customize their privacy settings. Moreover, one can easily identify that the two choices in \Cref{fig:cookiebannerb} are not equal: While the button that gives consent to cookies is highlighted by using a contrasting foreground color, the alternative is less prominent and is displayed using just a slightly different shade than the background color. This subtle difference in displaying alternatives can already be identified as a method to nudge the unassuming user towards accepting all available cookies, since the option to accept seems to be the default choice, while the alternative (customizing cookies) is illustrated as optional. Additionally, rejecting cookies is even harder for consumers, since the option for rejection is situated in a separate submenu, which subsequently requires significantly more clicks than to simply accept. One way to think of this user manipulation is in the terms of \textit{Dark Patterns}, a term coined by {\protect\NoHyper\citeauthor{Brignull2010}\protect\endNoHyper} in $2010$~\cite{Brignull2010}, and who described them as "tricks used in [web sites] and apps that make you do things that you didn't mean to", in this case accept to a variety of tracking cookies most users are not interested in. Using a more recent taxonomy by {\protect\NoHyper\citeauthor{Gray2018}\protect\endNoHyper}, this can be interpreted as an instance of interface interference, particularly a case of aesthetic manipulation~\cite{Gray2018}, since the user is nudged towards a decision simply by presenting two (or more) options as visually different, and thereby presenting one as the default. Moreover, this can be seen as a case of forced action, because usually the user has to interact with the cookie banner and needs to take an action before they can enter the web site. In this case, it can be classified as a version of "Privacy Zuckering", a pattern that tricks users into sharing more information about themselves than they intended. 

Besides the visual aspect, one can notice a second strategy in the above example, namely that the accept option is framed positively by stating "Yes, I'm happy", while the more neutral option leading to the privacy customization page is framed in a neutral tone as well. 
Recent work by \citet{Utz2019} and \citet{Luguri2019} showed that such design choices highly influence the decision making of users, and hence, one can assume that designers include these patterns deliberately into their work. Providing such insights is an important task, and we highly appreciate the work that is done to establish best practices and guidelines for designers to follow, such that they can avoid these malicious patterns. However, since advertisement and the corresponding tracking of user behavior is a crucial income source for many vendors, it can be assumed that even with suitable guidelines at hand, some designers will implement dark patterns nonetheless. For that reason, our work focuses on the automatic detection of dark patterns on web pages, and in the following, we outline our current work and give an example in the context of cookie banners. Throughout this work, we stick to the taxonomy of \citet{Gray2018}, and do not formulate our own definition of dark patterns.

\section{Counteracting and Automatic Detection of Dark Patterns}
\label{sec:detection}
Establishing best practices for designers to follow, such that they can avoid dark patterns, is a valuable contribution, but does not address web designers who use dark patterns willfully in their work. The protection of consumers against malicious design patterns needs to incorporate other means as well. Firstly, it is important to inform users about the usage of such patterns, and counteracting this asymmetry of power by informing consumers is also part of a study conducted by the \citet{NCC2018}. Secondly, researchers like \citet{Luguri2019} or \citet{Weinzierl2020} evaluate the (un)lawfulness of dark patterns, and discuss in which ways legislators can enact restrictions to the usage of such malicious patterns. Thirdly, one possibility is to warn consumers when they encounter usages of dark patterns on the web. Our work approaches this last strategy, in which we aim to automatically detect the usage of patterns on web pages. To the best of our knowledge, there is not much work that was conducted in the general field of pattern detection on web pages. One exception to this is the work by {\protect\NoHyper\citeauthor{Mathur2019}\protect\endNoHyper} who used automatic crawling to detect dark patterns on shopping web sites. However, their focus was on the creation of a dark pattern taxonomy and to assess how common dark patterns are on the examined shopping web sites~\cite{Mathur2019}, while our goal is to build a general framework to detect dark patterns on arbitrary web pages.

The challenges in this context are manifold. To begin with, web pages are heterogeneous in nature. First, the content of web pages has a wide range of topics, e.g., news, games, or shopping. Second, the structure of web pages is heterogeneous as well, making it hard to design a general approach to detect certain patterns in a diverse set of pages. Therefore, it is mandatory to develop a framework to detect patterns on a wide range of web pages. This model does not need to be specific to dark patterns, but should at least be able to incorporate them easily. Taking the heterogeneity of web pages into account, it is clear that purely rule-based methods are not sufficient to tackle the problem of dark pattern detection. An additional problem is the rarity of occurrences of dark patterns on web pages in contrast to regular page elements or other more frequent patterns that have no malicious intent. This makes it mandatory to consider solutions from the field of rare pattern detection and unbalanced classification as well.

For these reasons, our efforts aim to solve the problem using machine learning approaches to detect dark patterns as generically as possible. The employed model is based on input features from the Document Object Model (DOM), a tree-like representation of an HTML document, whereas each node of the DOM represents a part of the web page's document. In this tree, each node can be treated as a potential candidate of representing a dark pattern, and the detection can now be framed as a classification problem in which the algorithm decides for each node if it is in fact a malicious pattern. To be more precise, one can (and should) also consider subtrees of the DOM as a potential pattern, but for the sake of simplicity we restrict the discussion to nodes in this work. Since single elements of a web page are rarely a dark pattern in isolation, one needs to also consider the neighborhood of elements, and most often a node can only be classified as malicious if multiple conditions are met. Therefore, graph neural networks offer a promising perspective, since the DOM is already a tree, and hence, a hierarchical graph structure, and this approach could take into account the neighborhood of a node as well. However, considering that most neural architectures for classification need large amounts of labeled data to be trained, the problem that dark patterns are in fact a rare phenomenon becomes especially problematic. One approach to this can be active learning, an annotation technique for machine learning problems that queries those elements to a user for which the classification algorithm is unsure and as a result only critical cases need to be manually labeled~\cite{Aggarwal2014, Settles2009}. This can enable researchers to reduce the time consuming manual labeling process, whereas only an initial seeding set of examples needs to be labeled, and during application time additional samples can be gathered.

Lastly, the outlined approach does not consider dynamic aspects of dark patterns. Often, dark patterns only emerge when the user interacts with a web page or in the context of multiple web pages that are visited in succession. Therefore, it is necessary to extend the model in ways that incorporate dynamic aspects, e.g., by treating the DOM as a dynamic graph. However, our current research does not consider this aspect yet, but aims to integrate it in future iterations.

\section{Preliminary Results}
\label{sec:results}
\begin{figure}[t]
  \centering
  \includegraphics[width=0.9\linewidth]{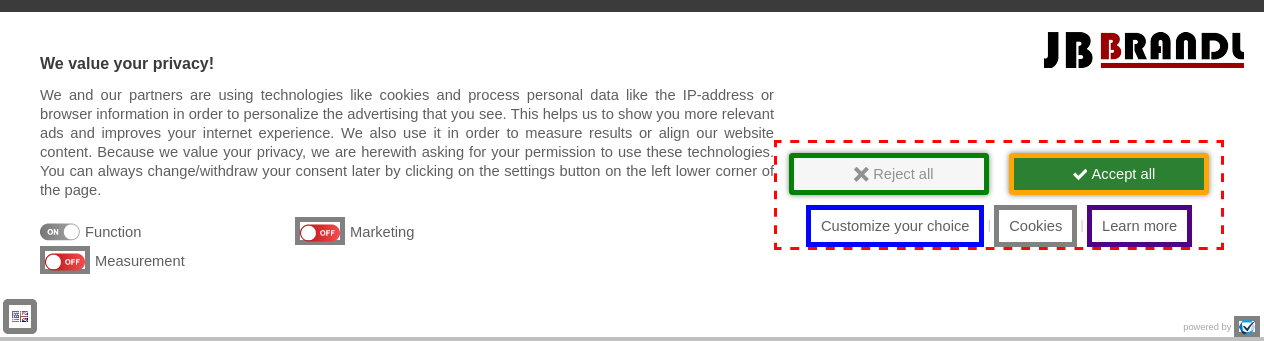}
  \caption{An automatically extracted cookie banner from \url{www.anhaenger-ersatzteile24.de/}. Colored borders are inserted via DOM manipulation of our program. Most importantly, an orange border indicates a detected "accept" and green a detected "reject" button. The dashed border indicates that the program has found an instance of visual interference in the cookie banner and is the lowest common ancestor of the "accept" and "reject" button.}
  \label{fig:example}
\end{figure}

Instead of directly approaching the more complex problem of generally detecting dark patterns, this section limits itself to a more confined task, the detection of problematic patterns within cookie banners. The algorithm employs a bottom-up approach, in which first low-level elements containing certain keywords are identified, and then larger segments of content around these low-level elements are iterated to find suitable cookie banner candidates. If multiple candidates are found, heuristics like the text length are used to determine the cookie banner.

By applying the implemented algorithm to more than $4000$ German web sites\footnote{Domains ending in \texttt{.de} and using the Python package polyglot for language detection of text content on the web page.} extracted from a list of the top one million web sites according to \url{Alexa.com}\footnote{\url{https://www.kaggle.com/cheedcheed/top1m?select=top-1m.csv} (accessed on $3$. February $2021$)}, around $2800$ cookie banners were extracted and analyzed. By utilizing features like the HTML tag of elements, it was possible to obtain a large amount of clickable elements within the banners. We then extracted textual features from those elements, and used clustering techniques to find different groups of buttons with regard to their textual content. Based on the initial clustering and a manual relabeling of critical items, a Support Vector Classifier is trained to distinguish between multiple button types. In this work, the most important distinction is between "accept" and "reject" buttons. By retrieving the CSS style information for each element, it is possible to identify if the two options ("accept" and "reject") are visualized similarly. In cases where the visual appearance is too dissimilar, site users can now be warned and it can be indicated that the acceptance of all cookies is in fact not the only option, even if the web site operator tries to suggest otherwise.

\Cref{fig:example} shows the outcome for a single web page, but nonetheless, the implemented framework is powerful enough to detect cookie banners on a wide range of web pages, and reliably identifies cases of aesthetic manipulation in the form of dissimilarly visualized user choices.

\section{Limitations}
\label{sec:limitations}
The automatic detection of dark patterns has also limitations that need to be acknowledged. First of all, not all types of dark patterns as classified by \citet{Gray2018} will be detectable by the proposed framework. For example, gamification is a problem that usually is not specific to web pages, but is part of apps that cannot directly be addressed by the model. Another instance is the "roach motel", a pattern that describes a situation a user can get into very easily but where it is hard to get out again, e.g., by being forced to send a physical letter to cancel a premium subscription. Oftentimes, it is not disclosed at subscription time that such action is necessary, and hence, it is hard to detect such an instance automatically.
Finally, the detection of dark patterns shares challenges with ad blockers. It is rarely possible to foresee which new patterns are going to emerge, and as a result, detection measures are always reactive, and rely on practitioners that constantly update the existing pattern databases as well as engaged consumers that point out new occurrences.

\begin{acks}
This work is part of the Dark Pattern Detection Project (dapde), which is funded by the German Federal Ministry of Justice and Consumer Protection.
\end{acks}

\bibliographystyle{ACM-Reference-Format}
\bibliography{dapda_paper.bib}


\end{document}